\documentclass[prd, twocolumn, nofootinbib, floatfix]{revtex4-1}

\usepackage{amsmath}
\usepackage{graphicx}
\usepackage{enumitem} 
\usepackage{dcolumn}
\usepackage{bm}
\usepackage{epsfig}
\usepackage{amssymb,latexsym,mathrsfs}
\usepackage{graphicx}
\usepackage{color}
\usepackage{hyperref}
\usepackage[strings]{underscore}
\hypersetup{
    colorlinks=true,
    linkcolor=red,
    citecolor=blue,
} 

\newcommand{\be}{\begin{equation}}
\newcommand{\ee}{\end{equation}}
\newcommand{\bs}{\begin{split}} 
\newcommand{\bea}{\begin{eqnarray}}
\newcommand{\eea}{\end{eqnarray}}

\newcommand{\om}{\Omega_m}

\newcommand{\gm}{G_{\rm matter}} 
 
\newcommand{\gl}{G_{\rm light}}

\newcommand{\al}{\alpha}

\newcommand{\almnow}{\alpha_{M,0}} 
 
\newcommand{\hiclass}{{\tt hi\_class}}

\begin{document}

\title{Gravity's Islands: Parametrizing Horndeski Stability} 

\author{Mikhail Denissenya${}^1$, Eric V.\ Linder${}^{1,2}$} 
\affiliation{${}^1$Energetic Cosmos Laboratory, Nazarbayev University, 
Astana, Kazakhstan 010000\\ 
${}^2$Berkeley Center for Cosmological Physics \& Berkeley Lab, 
University of California, Berkeley, CA 94720, USA}

\begin{abstract} 
Cosmic acceleration may be due to modified gravity, with effective field theory 
or property functions describing the theory. Connection to cosmological observations 
through practical parametrization of these functions is difficult and also faces the 
issue that not all assumed time dependence or parts of 
parameter space give a stable theory. We investigate the relation between 
parametrization and stability in Horndeski gravity, showing that the 
results are highly dependent on the function parametrization. This can cause 
misinterpretations of cosmological observations, hiding and even ruling out key 
theoretical signatures. We discuss approaches and constraints that can be placed 
on the property functions and scalar sound speed to preserve some observational 
properties, but find that parametrizations closest to the observations, e.g.\ in 
terms of the gravitational strengths, offer more robust physical interpretations. 
In addition we present an example of how future observations of the B-mode polarization 
of the cosmic microwave background from primordial gravitational waves can probe 
different aspects of gravity. 
\end{abstract}

\date{\today} 

\maketitle

\section{Introduction} 

Acceleration of the cosmic expansion is a signal of new physics: a cosmological 
constant vacuum energy, a new scalar field, or new laws of gravity. As we extend 
the standard model into new theories, we must ensure that the foundation is 
sound and internally consistent. In particular, the theory should be free of 
pathologies such as ghosts and instabilities. For modified gravity, there is a 
wide class within effective field theory, Horndeski gravity the most general scalar-tensor 
theory with second order equations of motions, that has four free 
functions of time in addition to the cosmic background expansion. These can also 
be viewed as four property functions, describing properties of the scalar and tensor 
sectors and their mixing \cite{Bellini:2014fua}. 

Parametrization of these functions in a physically meaningful way -- with a 
clear connection to observables and a sound theoretical foundation -- has been 
a challenging task fraught with pitfalls \cite{1512.06180,1607.03113}. 
(Also see, e.g., \cite{1705.01960,1505.00174,koyama} for some theory characteristics dealing 
with the field definitions rather than the property functions.) 
Here we examine this in terms of 
sensitivity and characteristics, concentrating on stability from the theoretical 
side, while also investigating the impact of very general 
observational considerations such as agreement with general relativity at early 
times and possessing characteristics consistent with the late time expansion 
history (e.g.\ a de Sitter limit). Recently, \cite{Kennedy:2018gtx} has proposed 
the interesting idea of using  
stability, in terms of the sound speed of scalar perturbations, as the quantity 
to parametrize and deriving the property function behavior from this. In our  
analysis of the function space, and its relation to stability, we can 
assess the utility and generality of that approach, in addition to elucidating the 
characteristics of the property function space. 

Furthermore, we explore the sensitivity to the parametrization used on the 
physical results and constraints. For example, \cite{Mueller:2016kpu} 
demonstrated that the strength of modified gravity constraints could vary by almost two orders 
of magnitude depending on time dependence and priors assumed. This is a key 
question for the utility and robustness of comparing theory quantities such as 
property functions or sound speed to observables such as growth and clustering 
of matter structure and light deflection (gravitational lensing). 

In Section~\ref{sec:scan} we scan through property function space and 
elucidate the relation between stability and functional parametrization, and also 
give an example of an observational effect by calculating the B-mode CMB polarization 
signature of the property functions. 
We discuss specific theories in Section~\ref{sec:cases} and compare to 
analytic stability results. Section~\ref{sec:difq} examines the approach of using 
an explicitly stable parametrization of sound speed to map out the stable 
regions of property function space. In Section~\ref{sec:obs} we discuss 
observationally related issues such as the implications for the modified 
Poisson equation gravitational strengths $\gm$ and $\gl$, and the impact 
of a general relativity past and de Sitter asymptotic future on acceptable 
parametrizations and stability. We conclude in Section~\ref{sec:concl}.

\section{Property Function Space} \label{sec:scan}

\subsection{Property Function Basics} 

The property function approach of \cite{Bellini:2014fua} is a form of effective 
field theory for the gravitational action. Within Horndeski gravity, the most 
general scalar-tensor theory giving second order field equations, this involves 
four functions of time: 
$\alpha_K$, $\alpha_B$, $\alpha_M$, and $\alpha_T$, in addition to the 
background expansion given by the Hubble parameter $H(a)$. 
One of the attractions of this approach is that each function describes a 
physical property or characteristic of theory -- respectively the structure 
of the kinetic term, the braiding of the scalar and tensor sectors, the 
running of the Planck mass, and the speed of gravitational wave propagation. 

By specifying the form and parameters of the time dependent $\alpha$ functions  one 
picks a particular theory of gravity. However, not every such theory may 
be sound: they may exhibit a gradient (Laplace) instability or may suffer from ghosts, rendering the theory unviable. Thus one must check the assumed 
parametrization of the property functions to assure the absence of such 
pathologies. 

The no ghost condition is easily tested, as it involves a simple combination of 
the kineticity $\al_K$ and braiding $\al_B$ functions, requiring the   condition 
\be 
\al\equiv \al_K+(3/2)\al_B^2 \ge 0 \label{eq:ghost} 
\ee 
be satisfied. As long as we choose $\al_K\ge0$ this will hold. This also automatically makes the denominator of the sound speed (see below) positive as well. 

To guarantee the theory of gravity is free of gradient instability, the form (and parameters) 
of the time evolution of property functions must ensure the positivity of the speed 
of sound. As we see below, it is natural to set $\alpha_T=0$ and so the  simple 
analytic expression for the speed of sound becomes 
\be 
c_s^2=\frac{(1-\al_B/2)(2\al_M+\al_B)+(\al_B/2)(\ln H^2)'+\al'_B}{\al_K+(3/2)\al_B^2} \ , \label{eq:cs} 
\ee 
where a prime denotes $d/d\ln a$. The stability condition is then 
$c_s^2>0$, and as mentioned above, this becomes a condition on nonnegativity 
of the numerator. 

There are a number of publicly available Boltzmann codes, e.g.\  
\cite{Zumalacarregui:2016pph,Raveri:2014cka}, going beyond general 
relativity by implementing property functions. These can be used for 
calculating the cosmic microwave background (CMB) temperature and polarization 
power spectra and the growth of matter density perturbations and the matter power spectrum. 
They also test for stability and ghosts (though we can use the above 
analytic equations for this). However, the public versions of these codes 
are limited in the functional forms usable for the property functions 
(while user defined function modules will eventually be implemented in the public 
versions, they are not yet there in a robust 
state\footnote{Many 
thanks to code authors Marco Raveri and Miguel Zumalac{\'a}rregui 
for their clarification and help on this issue.}). 
We therefore use them with the implemented power law form of the time evolution. 

Testing stability is a critical first step in comparing theories to data, 
and extracting meaningful cosmological information. For this, we use the 
analytic expressions in Eq.~(\ref{eq:ghost}) and (\ref{eq:cs}), though we 
have tested our results against the Boltzmann code \hiclass. Of particular interest is the role 
of the property function parametrization assumed on which theories are 
allowed. That is, for what time dependence forms, and what values of parameters 
within the forms, do we select which parts of property function parameter 
space. More seriously, does the parametrization bias the physical 
interpretation, such as implicitly disfavoring standard 
theories such as $f(R)$ gravity?

\subsection{Checking Stability} 

Of the four property functions, the two with the greatest impact on cosmic 
survey observables are $\al_M$ and $\al_B$. The $\al_K$ function has 
minimal effect on subhorizon scales \cite{Bellini:2014fua} and we set it to a small 
value that does not affect the results (recall it does not enter into the 
numerator of Eq.~\ref{eq:cs}). 
The gravitational wave speed $c_T^2=1+\al_T$ has been tightly restricted to 
be close to one, i.e.\ $|\al_T|\lesssim 10^{-15}$, at present by gravitational 
wave and electromagnetic counterpart observations \cite{ligo}. While this 
does not guarantee $\al_T(a)=0$ for all times (see, e.g., \cite{1803.06368,1802.09447,derham}), that is the simplest case and we adopt $\al_T=0$. 

Thus we are interested in the $\al_M$--$\al_B$ space. For the power law 
time dependence we initially consider (see Secs.~\ref{sec:difq} and \ref{sec:obs} 
for other cases), $\al_i(a)=\al_{i,0}a^s$ where $a$ is the cosmic scale 
factor and a subscript 0 denotes the present value. The first important 
aspect to note is that restricting the parameter space, i.e.\ the 
amplitudes $\al_{i,0}$, too much can miss structure in the parameter space. 
Indeed, we will find ``islands'' appearing at larger $\al_{i,0}$ that 
might otherwise have not been found. Recall that the background expansion history, 
i.e.\ the Hubble parameter $H(a)$, is specified independently of the 
property functions. For concreteness and agreement with observations we take 
it to be given by the concordance flat $\Lambda$CDM 
cosmology, with present dimensionless matter density $\om=0.3$. For 
property function and some related studies 
away from $\Lambda$CDM, see for example \cite{Raveri:2014cka,Peirone:2017lgi,1703.05297}. 

\begin{figure}[htbp!]
 \centering
 \includegraphics[width=\columnwidth]{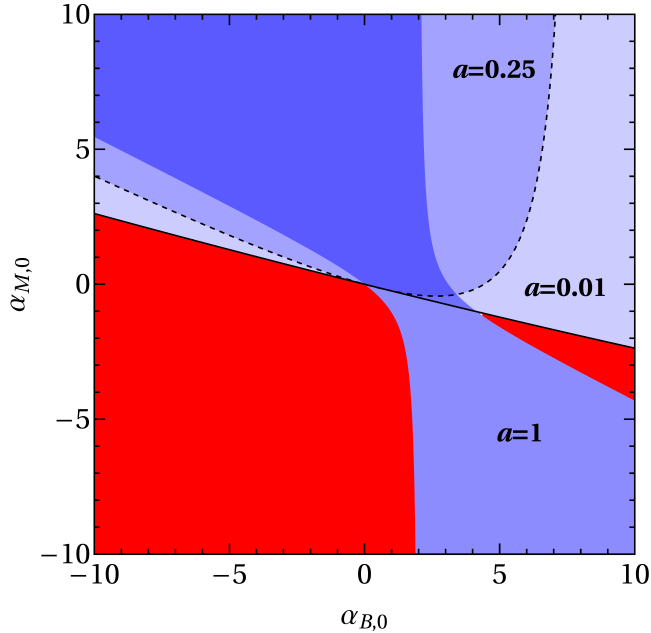} 
 \caption{Evolution of stability regions in the $\al_{B,0}$ and  $\al_{M,0}$ plane, 
for $\al_i=\al_{i,0}a^1$, obtained by varying the scale factor $a=0.01,0.25, 1$ 
are represented with different levels of blue color from light to dark. The dashed line denotes the boundary of the stability region at $a=0.25$ and solid black line shows the boundary between the upper stability and lower instability regions at $a=0.01$. The red 
region is unstable for all these scale factors. The intersection of the stable regions 
for all scale factors gives the viable parameter space.}  
 \label{fig:amaba} 
 \end{figure}

Figure~\ref{fig:amaba} shows how the shape of the stability region changes as we vary 
the scale factor at which the gradient stability is evaluated, here for $s=1$. 
At $a=1$ it extends into both 
left upper and right lower parts of the adopted parameter range with each part pinching 
in close to the origin in an hourglass shape. Smaller values of $a$ cut out most of 
the lower region, with intermediate values of the scale factor further diminishing this 
only slightly. The overall stability of the theory is determined by the intersection 
of the stable regions for all scale factors under consideration.

 \begin{figure*}[htbp!]
 \includegraphics[width=\columnwidth]{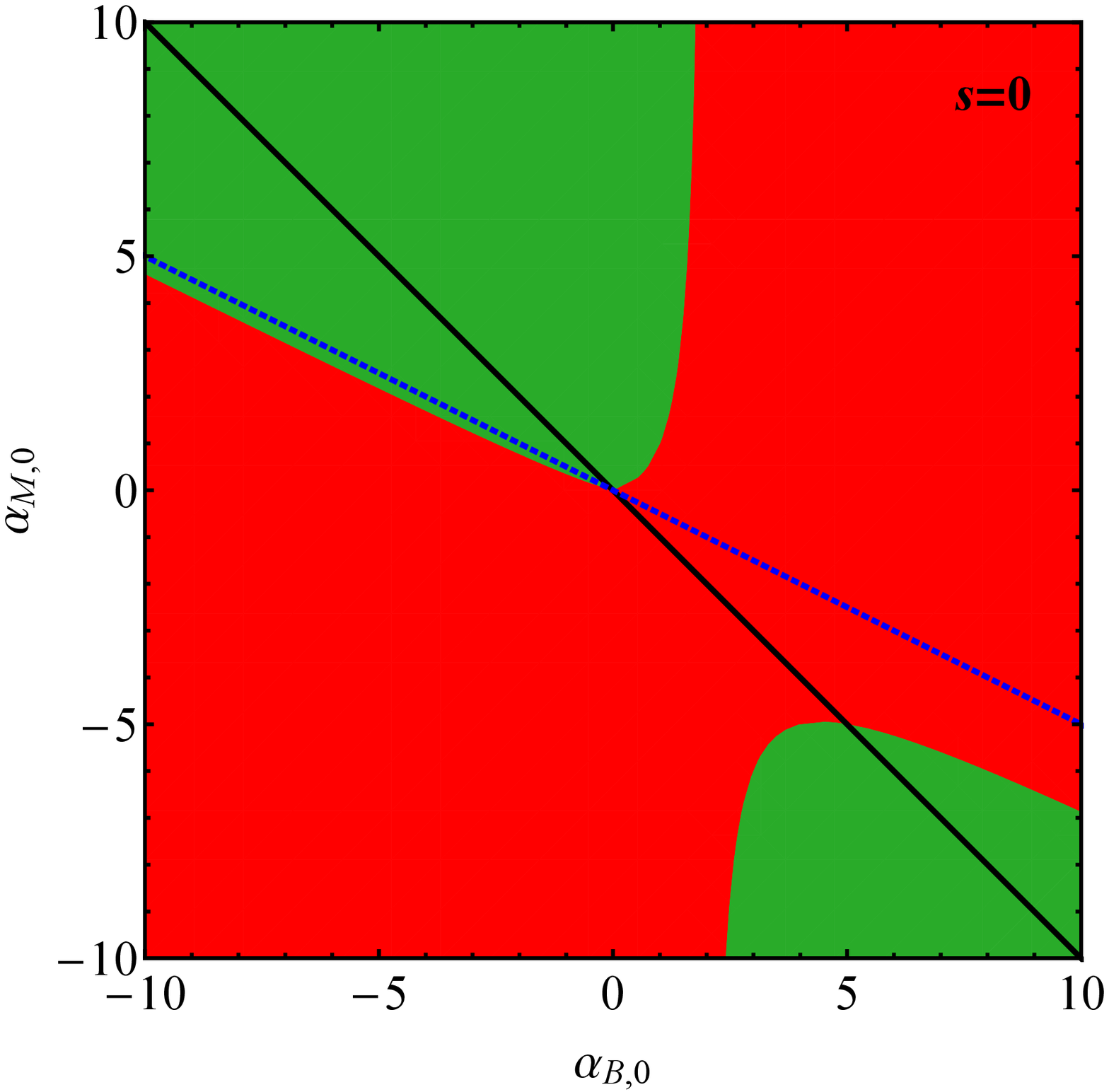} 
 ~ 
 \includegraphics[width=\columnwidth]{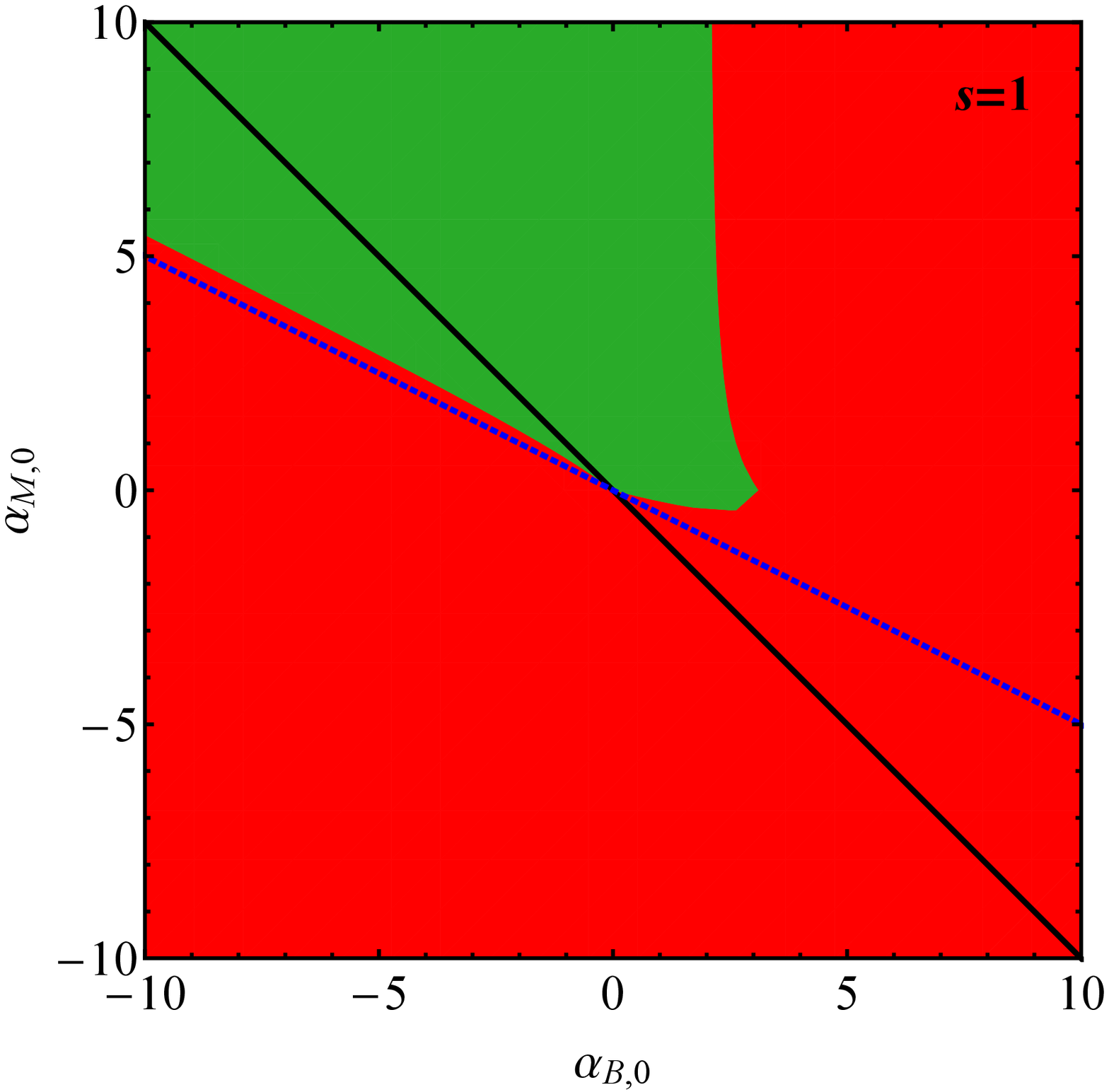}\\
  \includegraphics[width=\columnwidth]{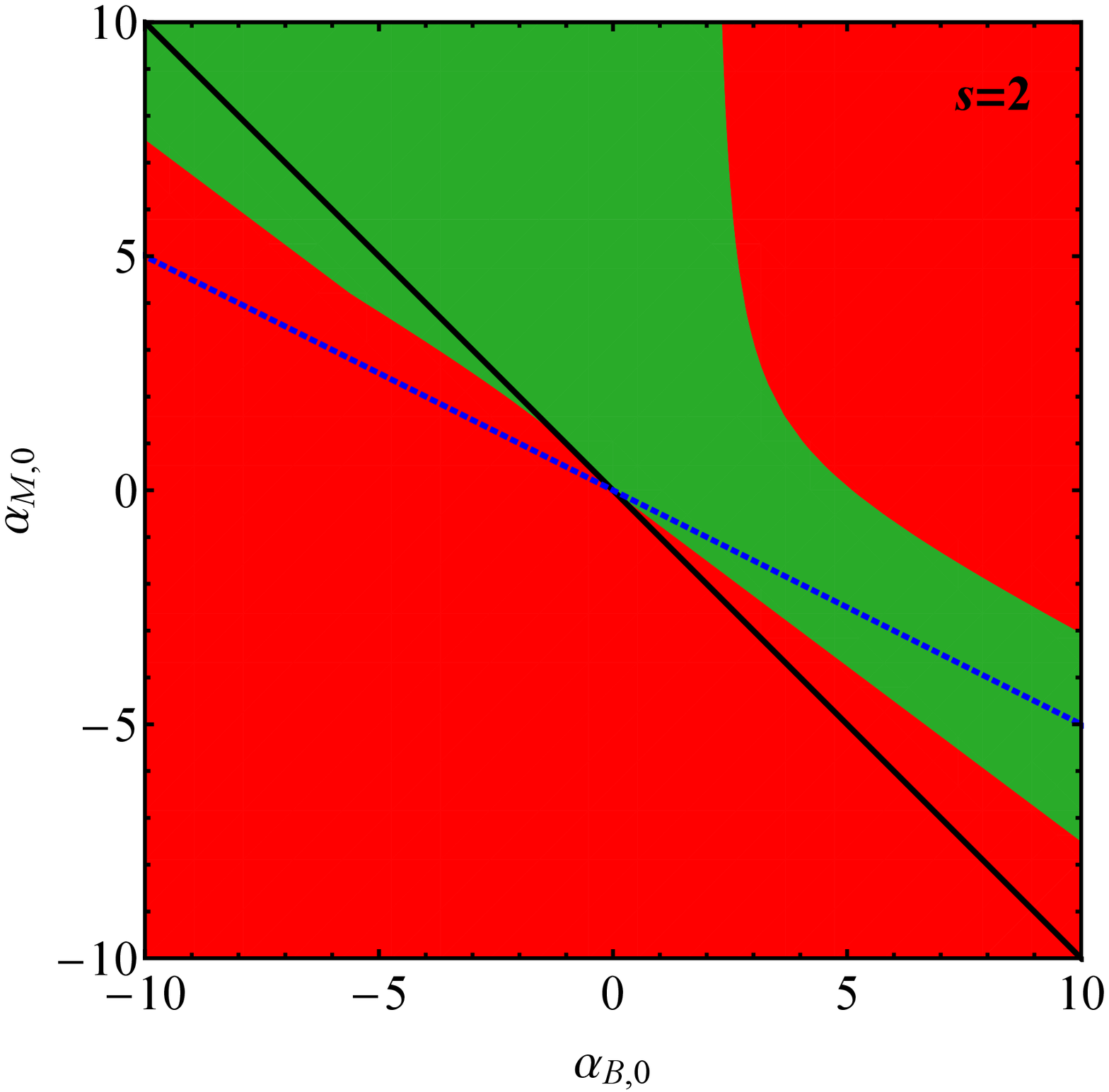} 
 ~ 
 \includegraphics[width=\columnwidth]{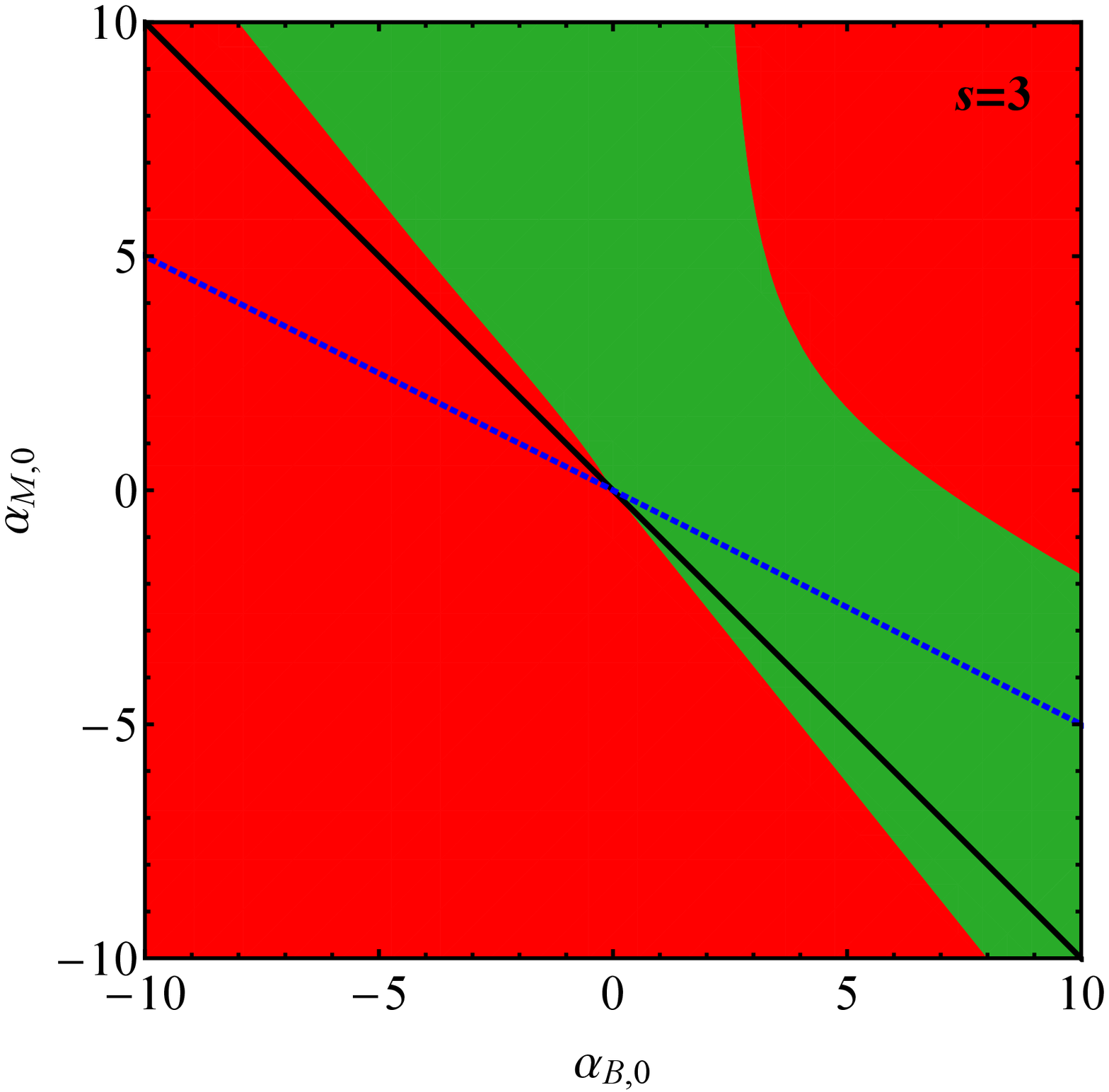}

 \caption{Regions of stability (green) and gradient instability (red) plotted in the $\al_{B,0}$ and  $\al_{M,0}$ plane determined over the range of scale factors $a=[0.001,1]$ for $\al_i=\al_{i,0}a^s$ and $s=0,1,2,3$. Black solid line corresponds to $f(R)$ theories ($\al_B=-\al_M$), blue dotted line corresponds to No Slip Gravity  ($\al_B=-2\,\al_M$).} 
 \label{fig:s02} 
 \end{figure*}

Figure~\ref{fig:s02} gives a clear illustration of how the viability of a model 
depends on its parametrization. Even within the family of power law scale factor 
dependence $a^s$, the regions of stability exhibit different geometries. The time 
independent case ($s=0$) shows disjoint islands of stability, while $s=1$ is mostly 
restricted to positive $\almnow$ and negative $\al_{B,0}$. As $s$ increases further, 
a tail develops to negative $\almnow$ and positive $\al_{B,0}$, thickening for 
larger $s$. At the same time, the stability region rotates as a result of the different weightings at different scale factors 
(cf.\ Fig.~\ref{fig:amaba}), lifting off the positive 
$\almnow$ cases of No Slip Gravity and then $f(R)$ gravity, while beginning to 
overlap the negative $\almnow$ cases of each of these in turn. Thus, the physical 
results one obtains in fitting data to theory are decidedly dependent on the 
property function parametrization used. This casts doubt on the utility of 
parametrization starting from the theory end, and adds support for parametrization 
starting from the observation end, as we discuss this further in Sec.~\ref{sec:concl}.  

\begin{figure}[htbp!]
 \centering
 \includegraphics[width=\columnwidth]{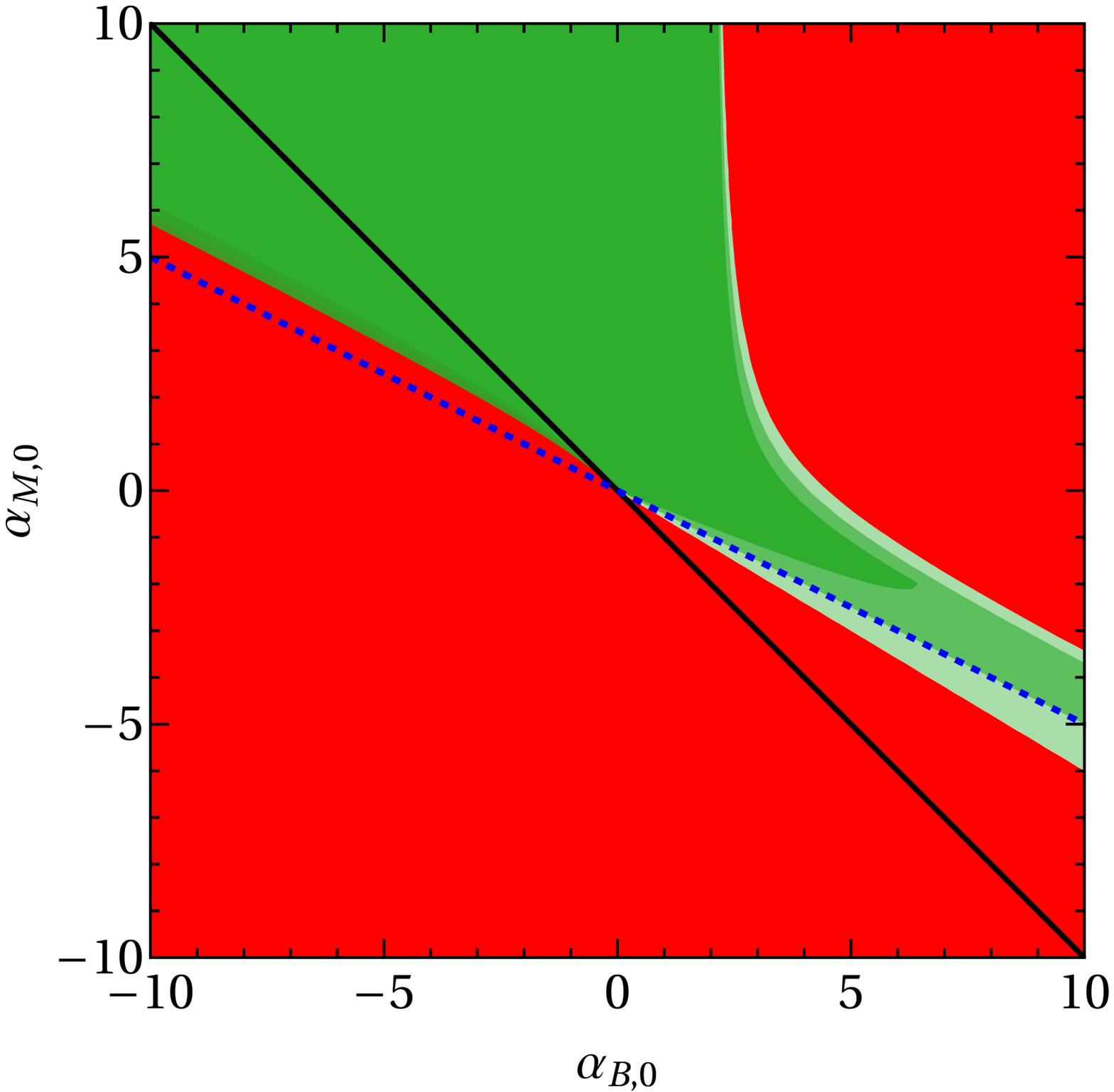}
 \caption{Regions of stability (levels of green) and gradient instability (red) plotted in the $\al_{B,0}$ and  $\al_{M,0}$ plane for $s=1.3$ (dark green), $s=1.5$ (green) and $s=1.7$ (light green). Black solid line corresponds to $f(R)$ theories ($\al_B=-\al_M$), blue dotted line corresponds to No Slip Gravity  ($\al_B=-2\,\al_M$).} 
 \label{fig:stabi} 
 \end{figure}

The extension of stability to the opposite quadrant, i.e.\ the ``tail'' to 
negative $\almnow$ and positive $\al_{B,0}$ is interesting to study further. One 
can show analytically that this occurs for $s=1.5$ (actually $s=3(1+w_b)/2$ for a 
background dominated by a matter component with equation of state $w_b$). We show 
the development of this tail in Fig.~\ref{fig:stabi} as $s$ goes from just below 
$s=1.5$ to just above. Not only does the tail extend to arbitrarily large values 
of $\al_{B,0}$ for $s\ge1.5$, but it does so along the No Slip Gravity line, 
$\al_{B,0}=-2\almnow$. Again, this follows analytically from Eq.~(\ref{eq:cs}) for 
the sound speed, since for large $\al_B$ it is the vanishing of the $2\al_M+\al_B$ term 
that prevents $c_s^2$ from going negative. Thus in some sense No Slip Gravity maximizes 
stability.

\subsection{CMB B-modes and $\al_M$} 

While the background expansion affects distance observables, and enters into 
growth of structure, the property functions affect perturbations in density 
and velocity, impacting growth of structure and gravitational lensing. However, 
they do not only affect scalar observables such as density perturbations. 
The propagation of tensor perturbations -- gravitational waves -- is affected 
by $\alpha_T$, which would modify the speed of propagation, and $\alpha_M$, 
which influences the friction term in the propagation equation and hence the 
evolving amplitude of the gravitational wave. As stated above we set $\al_T=0$, 
but it is interesting to examine the influence of $\al_M$ on gravitational 
wave observables. 

Since $\al_M$ also affects growth of density perturbations leading to cosmic 
structure, there is a close connection when $\al_T=0$ between the deviation of gravitational 
wave propagation from general relativity (in particular the distance to the 
source of gravitational waves compared to its counterpart electromagnetic 
distance) and the deviation of growth of 
structure from general relativity, as first explicitly highlighted in \cite{nsg}. Here, 
however, we explore primordial gravitational waves evidenced in cosmic 
microwave background (CMB) polarization B-modes. 

B-mode polarization arises from two contributions, the primordial tensor 
perturbations on large angular scales and the late time gravitational lensing 
conversion of E-mode polarization into B-modes on small angular scales. Since 
the lensing arises from structure in the 
universe it will be affected by growth deviations induced by $\al_M$ and 
$\al_B$. However the primordial B-modes will predominantly have the effect of 
an amplitude change due to $\al_M$. 

We use \hiclass\ to calculate the B-mode power spectrum, as well as the lensing 
deflection power spectrum. For the time dependence of the property functions we use 
the ``propto_scale'' option in \hiclass\ (see Table~1 of \cite{Zumalacarregui:2016pph}), 
so $\al_i=\al_{i,0} a^1$, i.e.\ $s=1$. 

\begin{figure}[htbp!] 
\includegraphics[width=\columnwidth]{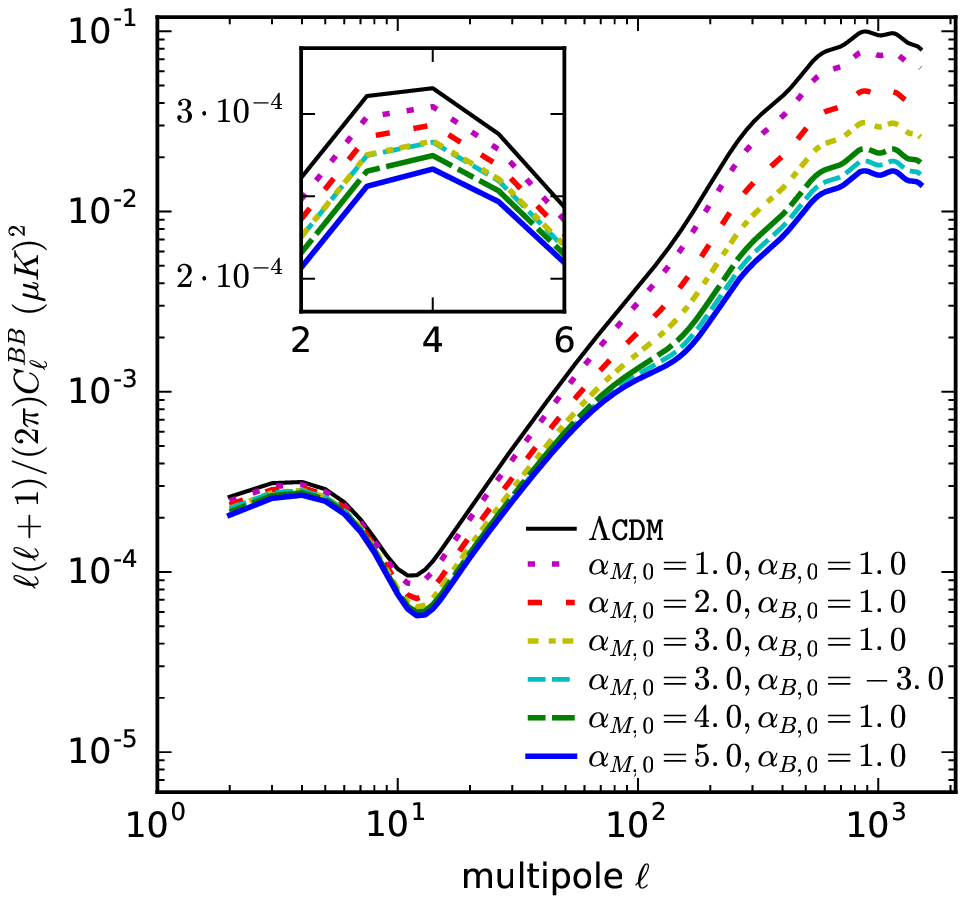} 
\caption{The primordial B-mode spectrum calculated using the property function parametrization of Horndeski models within the \hiclass, with time dependence 
$a^1$, for five values of $\alpha_{M,0}=1,2,3,4,5$, and $\alpha_{B,0}=1$ or $-3$, 
$\alpha_{K,0}=0.001$. The inset zooms in on the low multipoles, showing that only 
$\al_M$ matters. 
The tensor-to-scalar ratio $r=0.01$ and all spectra include the 
effects of gravitational lensing. The $\Lambda$CDM primordial spectrum is given by the 
solid black curve. 
} 
\label{fig:BBsp} 
\end{figure}

Figure~\ref{fig:BBsp} shows the CMB B-mode polarization power spectrum for 
several values of $\almnow$ and $\al_{B,0}$ in the stability region of the $s=1$ 
panel of Fig.~\ref{fig:s02}. The low multipole $\ell\lesssim10$ (large angular scale) 
reionization bump is due to primordial gravitational waves (here the inflationary 
tensor to scalar power ratio is taken to be $r=0.01$) and the high $\ell$ bump 
peaking at $\ell\approx1000$ is due to lensing.

The major effect of $\al_M$ is indeed a shift in amplitude (also see inset). 
Since $\al_M>0$ increases the friction term in the gravitational wave 
propagation, it decreases the gravitational wave amplitude, and hence B-mode 
power. 
Since $\al_B$ does not affect the gravitational wave propagation it leaves 
unchanged the B-modes at $\ell\lesssim10$, but it does enter into the growth of 
scalar density perturbations responsible for lensing at higher multipoles.

\section{No Slip Gravity and $f(R)$ Gravity} \label{sec:cases} 

To understand better, and confirm, the numerical results on stability we consider 
two specific theories of modified gravity. These will present one dimensional 
cuts through the $\al_M$--$\al_B$ space. One is $f(R)$ gravity, which 
imposes the relation $\al_B=-\al_M$, and the other is No Slip Gravity 
\cite{nsg}, with the relation $\al_B=-2\,\al_M$. For each of these we need 
parametrize only one function, which we take to be $\al_M(a)$. 

Proceeding along the lines of the previous section, we here adopt 
\be 
\al_M=\almnow\,a^s \ . 
\ee 
(The next sections consider further forms.) 
The analysis is particularly simple for No Slip Gravity as there the 
stability condition is simply 
\be 
(H\al_M)\,\dot{}\le 0 \qquad {\rm or} \qquad \frac{d(H\al_M)}{da}\le 0 \ . 
\ee 
This then becomes 
\be 
\almnow\,\left[(2s-3)\om a^{-3}+2s(1-\om)\right]\le 0 \ , 
\ee 
where we ignore radiation. We can readily define three cases: 
\begin{enumerate}[label=N\arabic*.] 
\item $s>3/2$: Stable for $\almnow<0$. 
\item $s<3\om/2$: Stable for $\almnow>0$. 
\item $3\om/2<s<3/2$: Unstable at some point in $a=[0,1]$. 
\end{enumerate} 
This agrees with the dotted line in Fig.~\ref{fig:s02} representing 
the No Slip Gravity condition $\al_B=-2\al_M$ (note $\almnow=0$ 
is just general relativity). 

For $f(R)$ gravity the stability condition in the power law $\al_M(a)$ 
model reads 
\be 
\almnow\,\left[1-s+\frac{\almnow a^s}{2}+\frac{3}{2}\frac{\om a^{-3}}{\om a^{-3}+1-\om}\right]\ge 0 \ . 
\ee 
This gives four cases: 
\begin{enumerate}[label=F\arabic*.]
\item $s>5/2$: Stable for $\almnow<0$. \label{list:f1} 
\item $0<s<1+3\om/2$: Stable for $\almnow>0$. \label{list:f2} 
\item $1+3\om/2<s<5/2$: Necessary but not sufficient condition for stability is $\almnow>2[s-(1+3\om/2)]$. \label{list:f3}
\item $s=0$: Stable for $\almnow>0$ and $\almnow<-5$. \label{list:f4} 
\end{enumerate}
This agrees with the solid line in Fig.~\ref{fig:s02} representing
the $f(R)$ gravity condition $\al_B=-\al_M$. (Note that $s=2$ requires 
$\almnow>1.11$; the exact stability condition for case \ref{list:f3}\ is analytic 
but messy, so we only show the simpler necessary condition.) For $s=0$ we see islands of stability 
appear that are disconnected from each other. This is an interesting 
property that we revisit in the next section when considering implicitly 
stable numerical parametrizations. 

There is physical motivation for these two theories, while there is not in 
general for ones with arbitrary $\al_B=-r\al_M$. However, we can use such 
a relation to show that: 
\begin{enumerate}[label=R\arabic*.] 
\item $s>3/2$: Stable for $\almnow>0$ when $r<4/(2s-1)$, for $\almnow<0$ 
when $4/(2s-1)<r<2$. 
\item $s<3/2$: Stable for $\almnow>0$ when $r<2/(1+s-3\om/2)$, 
unstable for $\almnow<0$. 
\item $r<0$: Unstable. 
\end{enumerate} 
It is interesting to note that $\al_B=-2\al_M$, i.e.\ No Slip Gravity, 
is a bounding model in the first case above. 

\begin{figure}[htbp!]
\includegraphics[width=\columnwidth]{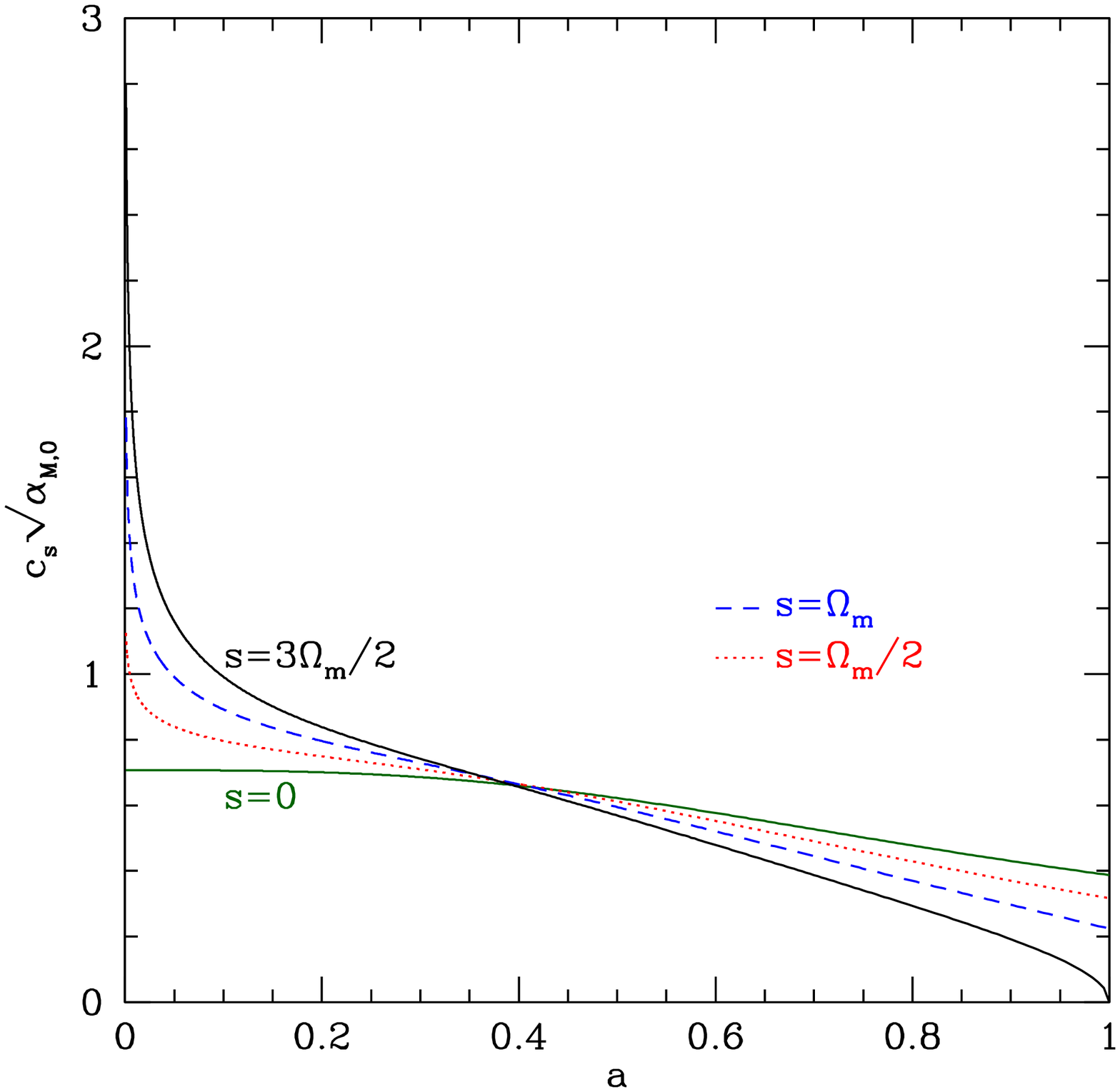} 
\caption{
The sound speed $c_s(a)$ is plotted vs scale factor for four cases 
of No Slip Gravity with $\al_M=\almnow a^s$ and $\almnow>0$. The 
upper limit of stability is $s=3\om/2$. We take $\om=0.3$. 
} 
\label{fig:cslowsnsg} 
\end{figure}

\begin{figure}[htbp!]
\includegraphics[width=\columnwidth]{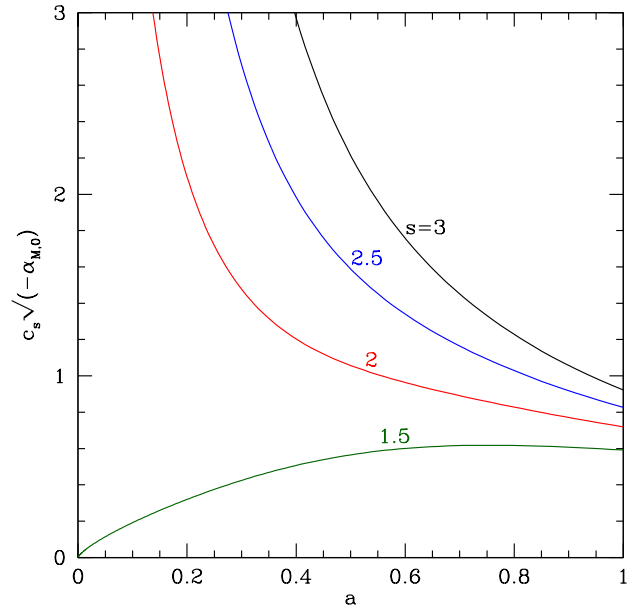} 
\caption{
The sound speed $c_s(a)$ is plotted vs scale factor for four cases 
of No Slip Gravity with $\al_M=\almnow a^s$ and $\almnow<0$. The 
lower limit of stability is $s=3/2$. 
} 
\label{fig:cshisnsg} 
\end{figure}

For the two physical theories we now consider the forms of the sound 
speed $c_s$ that these stable solutions represent. Figure~\ref{fig:cslowsnsg} 
and Figure~\ref{fig:cshisnsg} show $c_s(a)$ for various stable power law 
forms of No Slip Gravity, for $\almnow>0$ and $\almnow<0$ respectively. 
Note that $c_s^2\propto 1/|\almnow|$ so all the curves simply scale by this 
relation. At high redshift, $a\ll1$, the sound speed increases as 
$c_s^2\sim a^{-s}$, except for the bounding stability case $s=3/2$ and of course $s=0$. 
(This holds when $\al_K\ll \al_B^2$. Otherwise, $c_s^2(a\ll1)$ 
becomes of order $\al_M/\al_K$, which may go to a constant in the early universe. 
If $\al_K/\al_M\lesssim0.1$ the figures we plot are only 
significantly affected for high $c_s$ outside the range shown, so for 
simplicity we keep $\al_K=0$. When $\al_K$ gives a qualitative difference 
we will discuss it.)

\begin{figure}[htbp!]
\includegraphics[width=\columnwidth]{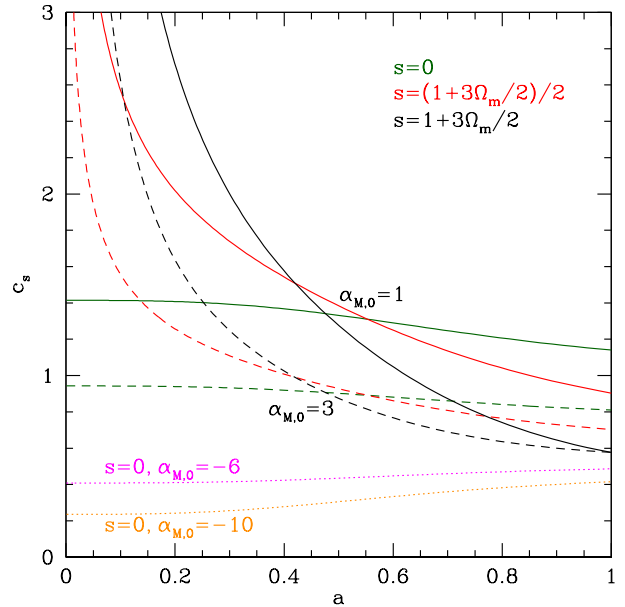} 
\caption{
The sound speed $c_s(a)$ is plotted vs scale factor for the two low $s$ cases 
of $f(R)$ gravity with $\al_M=\almnow a^s$. Note the $s=0$ case allows both positive 
and some negative values of $\almnow$. 
} 
\label{fig:cslowsfr} 
\end{figure}

For $f(R)$ gravity the sound speed behavior is qualitatively similar. 
Figure~\ref{fig:cslowsfr} illustrates the time dependence for the low 
$s$ stability range of cases \ref{list:f2}\ and \ref{list:f4}. Note the similar increase at high redshift. 
Now the sound speed does not scale simply with $\almnow$ so we plot 
two different values for each $s$.

\begin{figure}[htbp!]
\includegraphics[width=\columnwidth]{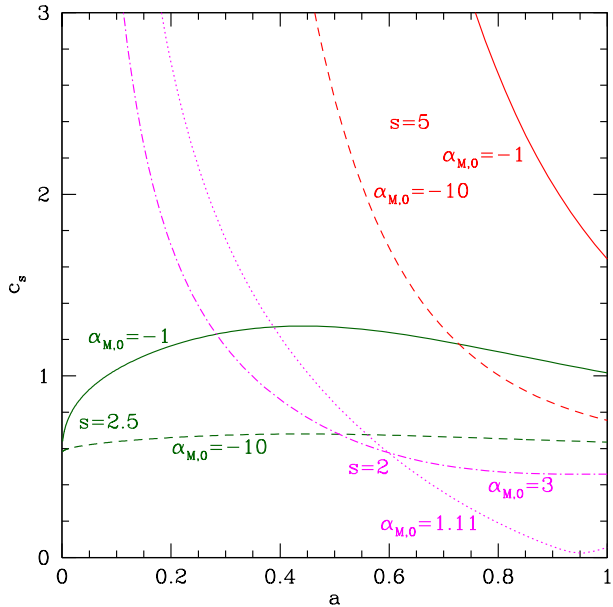} 
\caption{
The sound speed $c_s(a)$ is plotted vs scale factor for the two high $s$ cases 
of $f(R)$ gravity with $\al_M=\almnow a^s$. 
} 
\label{fig:cshisfr} 
\end{figure}

The higher $s$ stability ranges of $f(R)$ cases \ref{list:f1}\  and 
\ref{list:f3}\ are shown in Figure~\ref{fig:cshisfr}. 
The behavior is qualitatively similar, where $c_s^2$ grows as $a^{-s}$ at high 
redshift, except for the bounding stability case of $s=5/2$.

\section{Starting from Stability} \label{sec:difq} 

Recently, \cite{Kennedy:2018gtx} proposed the intriguing idea 
that rather than scanning through the 
property function space to check stability conditions, one rewrite the conditions 
$\alpha=\al_K+(3/2)\al_B^2\ge0$ and $c_s^2\ge0$ as a differential equation for 
$\al_B$. That is, for any input $\al_M$ (or the Planck mass $M_\star$ directly) and $c_s^2>0$ one would 
obtain $\al_B$ such that the pair ($\al_M,\al_B$) described a stable theory. 
This is an attractive feature, and furthermore one might hope that one has better 
guidance on priors for $c_s$ (at least its magnitude if not time dependence) than for 
$\al_i$. 

The differential equation, whose solution $\al_B(a)$ gives a guaranteed stable 
system, is 
\be 
\al'_B=\alpha c_s^2-\left(1-\frac{\al_B}{2}\right)(2\al_M+\al_B)-\frac{\al_B}{2H^2}\left(H^2\right)' \ , \label{eq:difq}
\ee 
where a prime denotes $d/d\ln a$. Note that we use the convention of \cite{Bellini:2014fua}, 
which differs from \cite{Kennedy:2018gtx} by a factor $-2$ in $\al_B$. 
While \cite{Kennedy:2018gtx} defines an 
auxiliary variable $B$ such that $B'/B=1+\al_B$ to obtain a second order 
differential equation (mathematically guaranteeing a real solution), this does not seem to yield 
any practical advantage so we keep the first order differential equation. 

This approach requires parametrization of $\al_M$ (or $M^2_\star$), and adds parametrization 
of $\alpha$ and $c_s^2$, and an initial condition on $\al_B$ 
(vs parametrization of $\al_B$, and $\al_K$ if desired, in the standard approach). 
One might hope to have better intuition on a 
parametrization for $c_s^2$ than for $\al_i$, but it is not obvious exactly how this would follow 
from some physical motivation (other than $c_s^2\ge0$). 
Implementing this approach requires adding a differential equation (to determine $\al_B$) 
versus the standard algebraic check of the positivity of $c_s^2$, possibly adding computation time. From Figure~\ref{fig:s02} we see that the stability region in 
$\al_M$--$\al_B$ space is not a particularly small fraction of the whole area, 
so the standard algebraic stability check should not cost more than a factor of a few in a uniform scan. 

To test these effects, we track the computational time required by the two approaches. In the standard approach, 
we uniformly scan over $\almnow$ and $\al_{B,0}$, check stability at $10^4$ redshifts from the early to late 
universe, and calculate the time required to obtain 1000 
stable cases. In the stability approach we do not have to 
check stability -- it is guaranteed -- but we do have to 
solve the differential equation to determine $\al_B(a)$. 
We input $\al_M(a)$ and $c_s^2(a)$ then evaluate $\al_B(a)$ at $10^4$ redshifts, and change the 
amplitude of $c_s^2$ at the present to obtain 1000 cases, again calculating 
the computational time. (Note that to minimize time we have not added further 
parameters to describe $\al$, but keep it fixed, just as we do with $\al_K$ in the 
standard approach.) 

In both 
cases we take the input functions to vary as $a^s$, with 
$s=1$; this gives the greatest disadvantage to the standard 
approach, since from Fig.~\ref{fig:s02} we see the smallest 
area of the parameter space is stable, 22\%. Despite this, the 
computational efficiencies are not significantly different (to generate 
1000 cases we find the standard approach is 7\% quicker). 

There is another important aspect. 
While the stability approach has the desirable property that it 
is pure in obtaining stability, it is not complete in the following sense. 
For a given parametrization of $c_s^2$, $\alpha$, and $\al_M$ one obtains a determined 
$\al_B$; this $\al_B$ will generally not have a strict proportionality to the chosen $\al_M$, 
i.e.\ $\al_B=-r\al_M$, and so physical models that enforce this, such as $f(R)$ gravity and 
No Slip Gravity, may be left out for at least some choices of parameter space. 

We pursue the extent of such restrictions further in the next section where we consider 
observational implications for parametrizations.

\section{Observational Considerations} \label{sec:obs} 

Two observational and physical considerations that we may want to take into 
account are that at early times we want the predictions to match general relativity, 
due to its success for primordial nucleosynthesis and the cosmic microwave 
background, and that at late times we may want the possibility of a de Sitter state 
to match the assumed background expansion history. These have particular 
implications for the property functions. 

\begin{figure}[htbp!]
\includegraphics[width=\columnwidth]{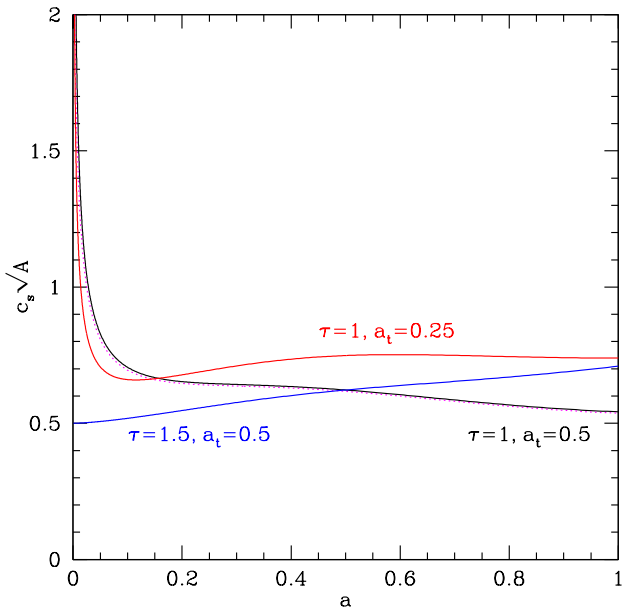} 
\caption{
The sound speed $c_s(a)$ is plotted vs scale factor for the ``hill'' form in No Slip Gravity, 
for different values of $\tau$ and $a_t$. The dotted, magenta curve uses the 
same parameters as the solid, black curve, but with $\al_K=0.1\al_M$; the 
function $\al_K$ has little effect within the region of interest. 
} 
\label{fig:csefoldnsg} 
\end{figure}

In general, we want the $\al_i$ property functions to vanish at early times 
to give general relativity in the early universe, and $\al_M$ to vanish at late times 
if we desire a de Sitter state (since it is a running of the Planck mass). The 
other property functions, and the sound speed, should go to constants in the de 
Sitter limit. Thus, rather than a power law form, these would have 
more of a ``hill'' form. To see what this implies for the sound 
speed, consider the $\al_M(a)$ hill form used for No Slip Gravity in \cite{nsg}, 
\bea 
\al_M(a)&=&A\,\left[1-\tanh^2\left(\frac{\tau}{2}\ln\frac{a}{a_t}\right)\right] \\ 
&=&4A\,\frac{(a/a_t)^\tau}{[(a/a_t)^\tau+1]^2} \ , 
\eea 
where $A$ is the amplitude, $\tau$ the steepness, and $a_t$ the location of the hill. This form (within No Slip Gravity)  
is guaranteed stable within the appropriate range ($\tau\le 3/2$). 
Figure~\ref{fig:csefoldnsg} illustrates the resulting $c_s(a)$.

We find that $c_s$ is fairly well behaved 
except at early times when it grows as $a^{-\tau/2}$. The boundary 
stability case of $\tau=3/2$ is an exception to this divergence; 
there $c_s(a\ll1)\to 1/(2\sqrt{A})$. Note that in general $c_s(a)\propto 
A^{-1/2}$, where $A$ is the amplitude $\al_M(a_t)$. The amplitude today is 
given by $\almnow=4A/(a_t^{\tau/2}+a_t^{-\tau/2})^2$, or $(8/9)A$ for 
$\tau=1$ and $a_t=0.5$. 

However, these early time behaviors assume $\al_K\ll\al_B^2$, so let 
us examine when this does not hold. The dotted, magenta curve in 
Fig.~\ref{fig:csefoldnsg} sets $\al_K=0.1\al_M$, so that it is not zero, and 
$\al_K\gg\al_B^2$ at early times. However, over late times relevant to 
observational tests of gravity, $\al_K$ makes little difference. 

That is the basic conclusion, but let's go into some further detail. As all the 
property functions become small in the approach to general relativity 
at early times, we can write 
\be 
c_s^2\to \frac{2\al_M+\al_B}{\al_K}+\frac{\al_B}{\al_K}\frac{d\ln\al_B}{d\ln a}\ . \label{eq:earlycs} 
\ee 
In many theories all the property functions will become proportional 
to each other (possibly with proportionality constant of zero) \cite{1512.06180,1607.03113}, 
and further evolve as power laws of the scale factor. In this case 
we see from Eq.~(\ref{eq:earlycs}) that at very early times the sound speed 
approaches a constant. Note that in No Slip Gravity the first term in Eq.~(\ref{eq:earlycs}) vanishes. If $\al_B\sim a^{3(1+w_b)}$ at early 
times, where $w_b$ is the background equation of state (e.g.\ 1/3 for 
radiation domination), then $c_s^2\to (3/2)(1+w_b)\al_B/\al_K$ for No 
Slip Gravity and $c_s^2\to (1/2)(1+3w_b)\al_B/\al_K$ for $f(R)$ gravity. 

Now considering late times, if the universe goes to a de Sitter state, then all 
time derivatives, e.g.\ $\dot H$ and $\al'_B$, vanish. This holds as 
well for $\al_M=d\ln M^2_\star/d\ln a\to0$. Thus we have 
\be 
c_s^2\to \frac{\al_B\,(1-\al_B/2)}{\al_K+(3/2)\al_B^2} \ . 
\ee 
All $\alpha_i$ should go to constants in the de Sitter state, and so 
the sound speed also goes to a constant. If the gravity theory has a 
relation $\al_B=-r\al_M$, then since $\al_M\to0$ then $\al_B$ also 
vanishes and $c_s\to0$. In particular, this holds for No Slip Gravity 
and $f(R)$ gravity. 

\begin{figure}[htbp!]
\includegraphics[width=\columnwidth]{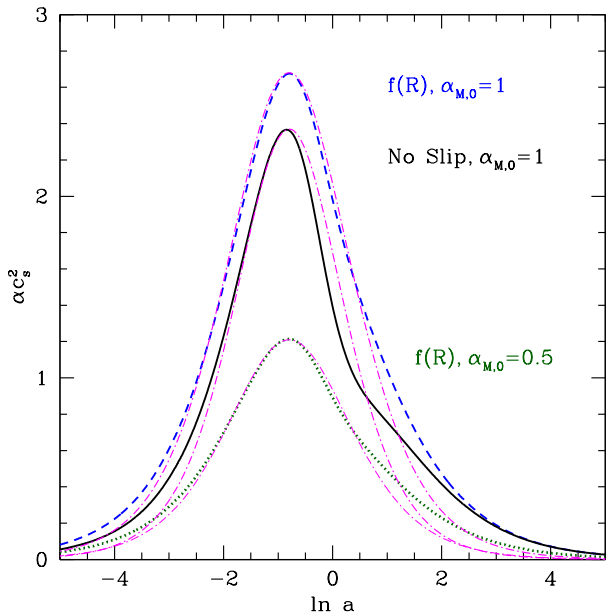} 
\caption{
The combined sound speed $q\equiv \al c^2_s(a)$ is plotted vs log scale factor for $f(R)$ gravity (dashed, blue curve for $\almnow=1$ and dotted, dark green 
curve for $\almnow=0.5$) and No Slip 
Gravity (solid, black curve) with $\al_M(a)$ given by the hill form. The dot 
dashed, magenta curves for each case give a fit to $q(a)$, using a similar 
hill functional form. 
Note for No Slip Gravity that $q$ simply scales with $\almnow$, while there is mild additional dependence for $f(R)$ gravity. 
} 
\label{fig:acshill} 
\end{figure}

Returning to the stability approach and its required parametrization of the sound 
speed, 
let us consider the two theories of $f(R)$ and 
No Slip Gravity to see what are the forms of $c_s^2(a)$  associated with them. This will give 
an idea for how 
straightforward it might be to start with a parametrization of sound speed. 
We can avoid the issue of $\al_K$ within the differential equation method by parametrizing the 
combination $q\equiv\alpha c_s^2$ which is all that enters, rather than $\alpha$ and 
$c_s^2$ separately. 
We still need to parametrize $\al_M$ or $M^2_\star$. 
For $f(R)$ we can evaluate $q$ using 
$\al_B=-\al_M$, and for No Slip Gravity using $\al_B=-2\al_M$; for both we use the hill 
form of $\al_M(a)$. Figure~\ref{fig:acshill} shows the derived $q(a)$.

These cases appear more tractable to parametrization than those from the 
previous power law cases. But that functional sensitivity means it is not clear that one can 
fruitfully employ one simple general form for $c_s^2(a)$ in the stability 
approach and capture variations in $\al_i$. Nevertheless, let us attempt to go 
one step further, parametrizing the derived $q(a)$ from a true input theory, 
as in Fig.~\ref{fig:acshill}, and seeing if the stability approach then 
accurately reconstructs the true theory. From Fig.~\ref{fig:acshill} 
a hill form, shown by the dot-dashed, magenta curves, appears a reasonable 
approximation to $q$, at least over the range of most observational 
interest$\,$\footnote{Note 
that such a parametrization adds three more parameters to the three from 
parametrizing $\al_M$ and one initial condition on $\al_B$. One $q$ parameter 
can actually be predicted based on the early time limit if one 
assumes all property functions are proportional there, but use of this value 
gives a poor fit, as does assuming parameters matching those of the input 
$\al_M$ form.}.  

\begin{figure}[htbp!]
\includegraphics[width=\columnwidth]{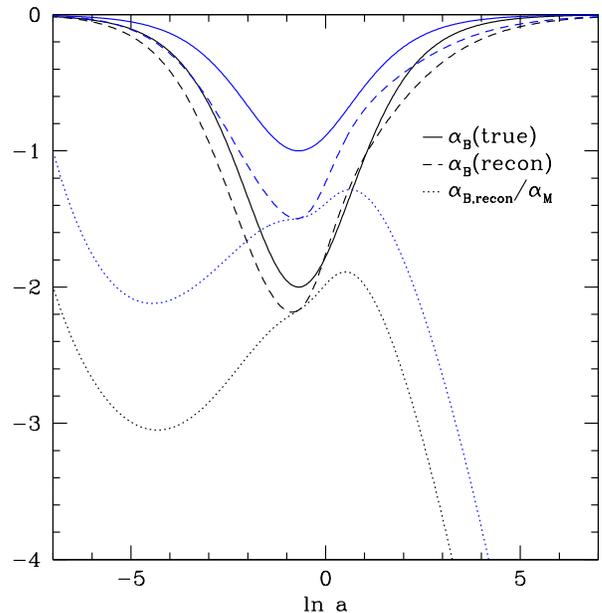} 
\caption{
Approximating the exact solution for $q$ by a hill form, i.e.\ a reasonable parametrization 
attempt, does not reconstruct accurately the input gravity theories. 
Solid curves are the true $\al_B$ for the input No Slip Gravity (dark black) or $f(R)$ gravity (light blue) theories, while dashed 
curves show the reconstruction based on using parameters that match 
the $\alpha c_s^2$ curves in Fig.~\ref{fig:acshill}. Dotted curves show the 
ratio $\al_{B,recon}/\al_{M,input}$; if the 
reconstruction were accurate then these curves should be horizontal at $-2$ for No Slip Gravity and at $-1$ for $f(R)$ gravity. 
While $\al_B$ may look 
of a similar hill form as $\al_M$, the No Slip Gravity relation $\al_B=-2\al_M$ or the $f(R)$ gravity relation $\al_B=-\al_M$ is not 
followed. 
} 
\label{fig:abcshillnsg} 
\end{figure}

We then use the same input function $\al_M(a)$ as in Fig.~\ref{fig:acshill}. 
Furthermore, we take initial conditions on $\al_B$  
such that it has the characteristic 
$\al_{B,i}=-2\al_{M,i}$ of a No Slip Gravity theory or $\al_{B,i}=-\al_{M,i}$ of 
$f(R)$ gravity. We use these as inputs to solving the 
differential equation for $\al_B(a)$ in the stability approach. Figure~\ref{fig:abcshillnsg} 
shows the results.

The reconstructed $\al_B(a)$ do not closely match the input truth. 
While $\al_B$ is roughly a hill form as it should be in mirroring $\al_M$, the 
amplitude and width are larger than they should be. 
Moreover, we do not recover the {\it class\/} of gravity theory, i.e.\ the 
ratio $\al_B/\al_M$ that are the characteristics of No Slip Gravity and $f(R)$ 
gravity. These key ratios are, in the reconstruction, neither constant nor centered on the right values for the two theories. 

Finally, if one propagates the reconstruction to the modified Poisson 
equation gravitational strengths, $\gm$ and $\gl$, one breaks 
characteristics such as $\gm=\gl$ for No Slip Gravity and also obtains 
pathological results at some redshifts as their denominators vanish 
due to inaccuracy of the reconstructed $\al_B$ and $\al'_B$. (See 
\cite{1712.00444} for a different study of the impact of stability on the 
gravitational strengths.) This is of particular concern since 
they are closely related to observables. It 
appears that even modestly inexact parametrization of the sound speed can lose 
significant information on the nature of modified gravity. 

If even these two viable theories, much less complicated than many Horndeski 
theories, cannot easily parametrize the 
essential element, $q$, entering the stability approach, and give 
rise to accurate physical interpretation, then the utility of 
property function (and sound speed) parametrization seems to lack robustness. We discuss an alternative in the Conclusions.

\section{Conclusions} \label{sec:concl} 

Modified gravity as an explanation for cosmic acceleration is a highly attractive concept, 
and has been connected to the observations in an increasingly sophisticated manner in recent 
years. If one wants to extract general physical characteristics of the theory, rather than 
working within one specific theory (with a particular functional form assumed, and particular 
values for the parameters assumed), then approaches such as effective field theory or property 
functions or modified Poisson equations are quite useful. 

However, these all contain functions that themselves need to be parametrized. Even before 
engaging in detailed calculations of such parametrized theories one must check that the theory 
is sound: lacking ghosts and instability. We examined in some detail the relation between the 
functional parametrization in the property function approach and the stability of the theory: 
the relation is not trivial. In particular, we showed how the stability evolves with redshift, 
picking out different regions of parameter space that can have complex structure (see 
Fig.~\ref{fig:amaba}). The final allowable stable part of parameter space is the intersection 
of stability for all redshifts. This can exhibit disconnected islands and also shows 
significant sensitivity to the time dependent form assumed for the property functions, even 
for the case where only two property functions contribute. Such sensitivity raises questions 
about the utility of the property function (or EFT) approach to give robust, general 
conclusions about modified gravity. 

Exploring this further, we considered a power law time dependence and studied the change in 
stability region as a function of power law index $s$. We derived various analytic expressions 
for the stability conditions and related them to two modified gravity theories: $f(R)$ 
gravity and No Slip Gravity. No Slip Gravity has the interesting property that it is a 
bounding theory: no theory that lies beyond No Slip Gravity in the relation $\al_B=-r\al_M$, 
i.e.\ with $r>2$, is stable for $s>3/2$ for all $\almnow<0$. The property function 
$\al_M$ is particularly interesting since it affects gravitational wave propagation, as 
well as density perturbations. We exhibited its effect on CMB B-mode polarization 
from primordial gravitational waves (and late time lensing), illustrating how it scales 
the power (as it does for late universe gravitational waves as well). 

A derived property from the property functions is the sound speed of scalar perturbations. 
We examined the implications of various parametrizations of the property functions on the 
sound speed, finding a great diversity in its behaviors -- power law dependence giving large 
$c_s$ at early times, bounded but nonmonotonic variation, both concave and convex variation 
-- all within the stability criterion and coming from simple power law time dependence of 
the property functions. This is directly relevant to the attractive idea by 
\cite{Kennedy:2018gtx} that one could start with enforcing stability by choosing a positive 
sound speed and then deriving the form of the property function $\al_B(a)$ preserving 
stability. That is, since $c_s$ is a function of $\al_M$ and $\al_B$, one can choose any 
two and determine the third function. However, our finding that simple $\al_i$'s give 
complicated $c_s$ casts some doubt on the approach of parametrizing $c_s(a)$. 

To explore this, we chose several forms of $c_s(a)$ (and $\al_M(a)$) and calculated the 
resulting $\al_B(a)$. We found that even if we chose a form $c_s(a)$ close to that predicted 
from a full theory such as $f(R)$ or No Slip Gravity, the reconstructed $\al_B$ and overall 
modified gravity was not faithful to the original. It broke essential physical characteristics 
such as injecting slip into No Slip Gravity or breaking the relation $\al_B=-\al_M$ in 
$f(R)$ gravity. Moreover, this stability approach was pure but not complete -- it did indeed 
guarantee stability but it did not (with reasonable guesses for the parametrized function 
$c_s(a)$) generate standard theories such as $f(R)$ gravity. 

Another relevant question is whether this stability approach is efficient. Removing the need for a stability check in the Boltzmann code saves 
computational time, but adding an extra differential equation to solve (and possibly 
increasing the overall number of parameters because one may have to account for $\al$, or 
$\al_K$, while it can mostly be ignored in the standard 
approach) compared to the standard approach of parametrizing 
$\al_B$ and $\al_M$ can cost time. We checked this and found there was no significant time savings from the 
stability approach, even when it did not involve an increased number of parameters. 

Finally, we investigated the impact of observational constraints on allowable parametrizations. 
One would like to impose that general relativity is restored in the early universe, so all 
the $\al_i$ go to zero. We explored the resulting implications on the sound speed. Similarly, 
one might look for a de Sitter state in the asymptotic future, and we discussed its 
implications 
on the property functions and sound speed. A useful parametrization that encompasses both 
these conditions is the ``hill'' form, and we compute $c_s$ and $\al_B$ in this case. We 
motivated use of the combination $q=\al c_s^2$ which enters the equation for $\al_B$, and 
showed 
this can be reasonably fit by the hill form, and in turn the reconstructed $\al_B$ looks 
qualitatively, if not quantitatively, similar to the input truth. 

However, we demonstrated that even small inaccuracies in the reconstructed $\al_B$, from 
residuals of the parametrization of the sound speed, can give rise to significant physical flaws. 
The denominators of the gravitational strengths $\gm$ and $\gl$ can spuriously pass through 
zero, giving pathologies. Combined with the lack of fidelity in preserving physical 
characteristics of known theories such as $f(R)$ and No Slip Gravity, and indeed the 
difficulty including them using straightforward parametrizations of the sound speed, this means that 
parametrization in terms of property functions or EFT is highly nontrivial, notwithstanding 
stability considerations. 

Parametrizations from the theory side, while undeniably attractive, unfortunately are found  
to be subject to issues of functional sensitivity and lack of robustness. However, there is a 
reasonable solution by moving closer to the observables. The gravitational strengths 
$\gm$ and $\gl$ entering the modified Poisson equations, directly related to growth 
of matter structure and light deflection, have been demonstrated to give robust and highly 
accurate descriptions of the observables, as well as key indicators to theory characteristics 
\cite{Denissenya:2017thl,Denissenya:2017uuc}. Such simple, model independent parametrizations 
as binning in redshift of these functions can be a highly useful first step in uncovering 
signatures of modified gravity.

\acknowledgments 

EL thanks Alessandra Silvestri for useful discussions and Yashar Akrami and the Lorentz Institute 
for hospitality. This work is supported in part by the Energetic Cosmos Laboratory and by 
the U.S.\ Department of Energy, Office of Science, Office of High Energy 
Physics, under Award DE-SC-0007867 and contract no.\ DE-AC02-05CH11231.

 
\end{document}